# Twenty-five years of J-DSP Online Labs for Signal Processing Classes and Workforce Development Programs

Andreas S. Spanias
SenSIP Center, School of ECEE
Arizona State University
spanias@asu.edu

*Abstract*—This paper presents the history of the online simulation program Java-DSP (J-DSP) and the most recent function development and deployment. J-DSP was created to support online laboratories in DSP classes and was first deployed in our ASU DSP class in 2000. The development of the program and its extensions was supported by several NSF grants including CCLI and IUSE. The web-based software was developed by our team in Java and later transitioned to the more secure HTML5 environment. J-DSP supports laboratory exercises on: digital filters and their design, the FFT and its utility in spectral analysis, machine learning for signal classification, and more recently online simulations with the Quantum Fourier Transform. Throughout the J-DSP development and deployment of this tool and its associated laboratory exercises, we documented evaluations. Mobile versions of the program for iOS and Android were also developed. J-DSP is used to this day in several universities, and specific functions of the program have been used in NSF REU, IRES and RET workforce development and high school outreach.

## I. Introduction

The first version of Java-DSP (J-DSP) we developed in 1997 and presented by our team at IEEE ICASSP in 1998 [1]. At that time there were other efforts that focused on using Java for visualization of signal processing concepts [2,3]. J-DSP however, was unique in that it established an entire object oriented visual programming environment that enabled students to establish online signal processing simulations.

Inspiration for developing visual J-DSP programming came from early MS DOS based programs such as Simulink [4] in 1990 for controls systems and the DSP experiments by Kamas and Lee [5] which were supported by Burr Brown MS DOS tools. A distinct difference of J-DSP from these early visual DSP environments, is that J-DSP is completely web-based and runs on a web browser. Hence J-DSP is universally and freely accessible. J-DSP enabled interactive labs and simulations on digital filters, Fast Fourier transforms (FFTs), z-plane pole-zero representations and random signal manipulations. An NSF award [6] from the then CCLI division enabled our team at Arizona State University (ASU) to develop additional functions to support online laboratories in senior level elective courses including DSP, Controls and Communications. The introduction of J-DSP in the undergraduate 4-credit DSP class at ASU became part of the syllabus of this class with one of the credits assigned specifically to the labs. Laboratories were developed for experiments with the impulse and frequency responses of FIR and IIR systems, pole-zero z-domain analysis, filter design, windows and FFTs, and random signals. One of our objectives with these online labs was to enable distance learners to receive hands-on experiences [7]. An assessment of these initial DSP labs was reported in [8,9]. Around the same time, efforts were launched to create Java software to enable embedding web-based demonstrations in web pages [10] and exporting to MATLAB J-DSP simulations [11]. Dissemination [12,13] and assessments were implemented at several universities including UT Dallas, University of Central Florida, Johns Hopkins University, MIT, University of Washington-Bothell, Prairie View A&M University, Rose-Hulman Institute of Technology, University of Rhode Island, University of New Mexico, and University of Cyprus. Additional functions were developed to support outreach in high schools [14]. An overall description of J-DSP and all functions and assessments until 2005 was presented in [15]. The J-DSP visual web-based programming environment is shown in Figure 1.

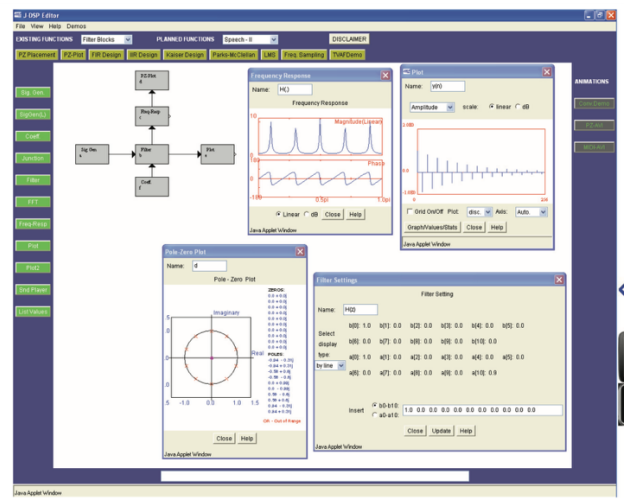

Figure 1. J-DSP simulation of a digital filter showing the functions engaged in the block diagram, the z-plane and frequency response plots and the impulse response graph.

Online laboratory exercises for classes were designed specifically for J-DSP. These included exercises on digital filters and their design, multirate signal processing, the Discrete and the Fast Fourier transform (DFT and FFT), random signal analysis, speech analysis-synthesis using linear prediction, and audio compression basics including examination of psychoacoustic models. The details of these online laboratory exercises and their evaluation are presented in the next section. More advanced J-DSP functions were developed for internal testing and graduate class deployment including adaptive filters [16-18], Hidden Markov models (HMMs) [19], computation of cepstral functions [20], and computation of Mel frequency cepstrum coefficients (MFCCs) [21].

In 2017, J-DSP went through a complete overhaul to enhance secure access. This was motivated by documentation of intrusions on Java applet software [22,23]. At about that time, publications appeared about transitioning education-oriented

Java applets to HTML5 [23]. Hence, we also launched the more secure HTML5 version of J-DSP [24] which also enabled advanced J-DSP functions for machine learning, speech recognition, and quantum Fourier transform (QFTs).

The rest of the paper is organized as follows. J-DSP online laboratories are described in Section II and HTML5 versions of the software are described in Section III. Machine learning and speech recognition experiments are described in Section IV and experiments with the QFT are described in Section V. In Section VI we describe mobile versions of J-DSP on iOS and Android platforms and Section VII describes the deployment of J-DSP in outreach and workforce development programs.

## II. J-DSP ONLINE LABORATORIES AND THEIR ASSESSMENT

The development of J-DSP online laboratories followed pretty much the sequence of topics we cover in our DSP class. These topics include discrete-time systems and FIR and IIR digital filters, transfer functions and z-transforms, FIR and IIR filter design, multirate signal processing, DFTs and FFTs and their properties, and discrete-time random signal manipulation. The J-DSP laboratory exercises are as follows:

- Lab 1 & 2; simulating digital filters and visualizing FIR and IIR impulse and frequency responses. In addition, z-plane and graphical placement of poles and zeros enabled students to examine stability and frequency response effects (Fig. 1).

- Lab 3; designing FIR and IIR filters using various methods. For example, for FIR linear phase design, students experiment with the Kaiser window, frequency sampling and Parks-McClellan methods. IIR design via analog filter approximations including Butterworth, Chebyshev and Elliptic are also examined (Fig. 2).

- Lab 4; The FFT and its properties were included in a lab where students experimented with spectral resolution and leakage using various windows. An exercise on peak-picking FFT spectra for speech compression was also assigned.

- Lab 5; A laboratory was designed for multirate signal processing enabling the students to experiment with downsampling and upsampling as well as QMF filter banks.

- Lab 6; Simulations with random signals examined autocorrelations, periodograms and parametric AR models were also enabled by J-DSP. In addition, linear prediction algorithms have been simulated for speech compression.

**Assessments;** All the laboratories have been individually assessed with pre- and post-quizzes on knowledge gained and student interviews on overall software usability. Details on these assessments are given in [15]. In Figure 3, the results of the pre-and post-quizzes are given for all the labs in [15]. In addition, we show a summary of quiz concepts for each J-DSP laboratory and the corresponding gains (from [15]). In all cases, we had substantial improvements especially in Labs 2 and 3 with 13.5% and 9.5% respectively. These gains are attributed mainly to the user-friendly interactive graphics which helped students better understand key concepts. These results along with the interviews provided feedback for the redesign of J-DSP blocks and inclusion of additional simulations with animations.

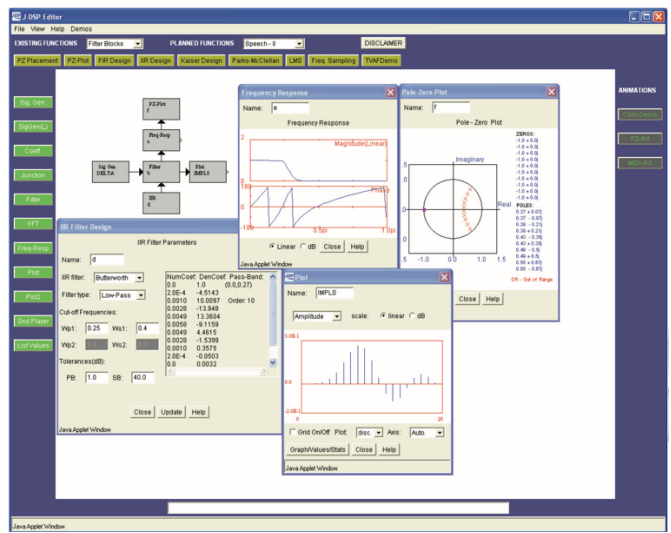

Figure 2. Butterworth filter design experiment using J-DSP with displays of the impulse and frequency responses, pole-zero plots and list of filter coefficients.

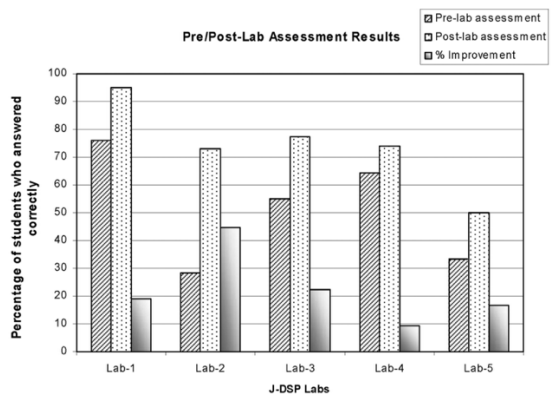

Figure 3. The pre- and post-lab assessments for all the exercises are shown along with the percentage of improvement (from [15]).

We note that J-DSP became particularly useful in delivering online lab experiences to remote learners during the COVID period (reported in IEEE FIE 2021).

## III. J-DSP TRANSITION TO HTML5

As noted before, Java applets posed security concerns in visualization examples [22-23] and in fact our J-DSP web site encountered intrusions. Most browsers at that time no longer supported Java applets and hence our team launched an effort [24] to re-design J-DSP for the HTML5 environment. The re-designed HTML5 J-DSP environment is fully browser-based with no external plugins and runs on all browsers. This secure version provided improved compatibility and a user-friendly graphical user interface (GUI) with no connectivity limitations.

The HTML5 version of J-DSP can process large data sets and support real-time data acquisition from mobile devices. The acquired speech data can then be used across different J-DSP functions to estimate spectra, compute autocorrelations and linear prediction (LP) parameters, examine quantization effects and even simulate an entire linear predictive coding (LPC) analysis-synthesis system. The new look of the J-DSP blocks in HTML5 is shown in Figure 4.

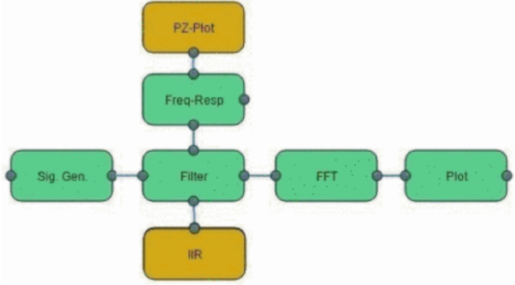

Figure 4. J-DSP simulation graphic in HTML5 [24] consisting of a digital filter, FFT, frequency response, and pole-zero plot.

## IV. J-DSP FOR MACHINE LEARNING AND SPEECH RECOGNITION APPLICATIONS

With the strong presence of machine learning (ML) and the emergence of new voice recognition models, we began developing exposition exercises that engaged FFTs, speech processing and AI. The HTML5 version of J-DSP enabled students to work with large datasets which in turn enabled training of ML models for speech recognition applications. In 2017, we introduced machine learning modules [25] in our undergraduate DSP course to cover speech and phoneme recognition. Students used existing J-DSP LPC functions and pole-zero computations to estimate the formants associated with phonemes. Additional functions were programmed in J-DSP for clustering data using the k-means algorithm [26, 27]. A speech recognition lab was designed to use speech data that is processed on a frame-by-frame basis. Windows are used and the LPC block computes autoregressive (AR) coefficients. Using this J-DSP simulation, students visualize the vocal tract spectral envelope and its resonant modes that enable them to compute the formants. k-means clustering is used to classify formants and recognize phonemes. Students experiment with the order of the LP algorithm, the number of clusters, and the frame length. Students also produce confusion matrices for the machine learning simulation and the classification of phonemes with and without noise. We also note that J-DSP was also used for gender classification in other universities [28].

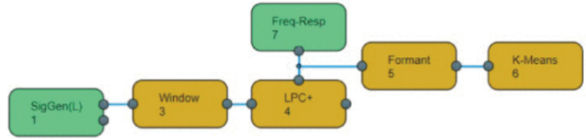

Figure 5. J-DSP simulation [25] that uses LPC for formant estimation and k-means clustering.

## V. QUANTUM FOURIER TRANSFORM FUNCTIONS AND J-DSP

One of our recent grants from NSF [29] focused on the development of quantum computing (QC) tools for exposition of undergraduate (UG) students and for the introduction of QC in our REU, IRES and RET programs. Our DSP class covers the FFT and its applications and one of the course modules addresses speech analysis-synthesis using Fourier transform peak picking [30]. Spectral component selection for speech analysis-synthesis based on perceptual criteria was also addressed in [31].

The development of J-DSP software for quantum Fourier transform simulations was presented in [32,33]. The theory of the QFT was developed in [34,35] and IBM Qiskit simulations were published in [36]. The J-DSP tasks on the development of QC tools consist of creating new functions that emulate the computation of the QFT and inverse QFT (IQFT). We note that the equation for the computation of the QFT has similarities with the equation of the DFT. The development of the J-DSP QFT tools started first by designing Qiskit circuits and an example for a 3-qubit simulation is shown in Figure 6.

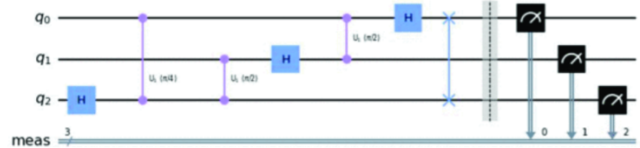

Figure 6. A 3-qubit QFT designed using Qiskit [32].

QFT and IQFT functions were developed in J-DSP for signal analysis and synthesis (Figure 7). The QFT function produces spectral components from an input signal. The signal input is represented in terms of a quantum state. The computation of the QFT components is based on a quantum measurement which is affected by quantum state measurement noise [37]. Noise models are used in J-DSP to emulate the effects of quantum noise in signal analysis-synthesis. Qiskit Aer has realistic noise models [38] to study QFT algorithm performance. By comparing ideal (noise-free) simulations with noisy simulations, students can directly observe the sensitivity of QFT-based signal analysis-synthesis to quantum noise. This provides a practical experience for introducing quantum noise and hardware constraints in undergraduate DSP laboratories. The J-DSP simulation diagram in Figure 7 shows the entire analysis-synthesis of a speech sequence using a QFT, an IQFT, peak-picking and noise emulation.

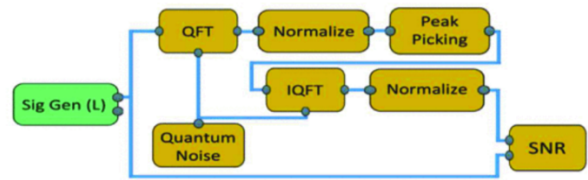

Figure 7. QFT-based signal analysis-synthesis [32]. The presentation of this tool received Top 3% recognition at ICASSP 2023 [53].

The QFT produces the frequency spectrum which is normalized. Peak picking then selects a compressed representation of the spectrum and the IQFT is used to reconstruct the signal. The quality of signal reconstruction is evaluated using the SNR function. Quantum noise can be introduced to evaluate noise effects on signal reconstruction. Students can vary the number of qubits, number of spectral peaks retained and the level of quantum noise. Additional details on results and assessments of the software and the associated J-DSP exercise are presented in [33]. A supporting module is given to introduce basic QC concepts to UG DSP, NSF REU and IRES students as well as NSF RET participants.

## VI. MOBILE VERSIONS OF J-DSP IN iOS AND ANDROID

With the emergence of iPhone and Android mobile phones and tablets our team began porting J-DSP on mobile platforms [39]. The use of mobile devices offers strong potential for engaging students due to their accessibility and interactive interfaces. Prior studies have shown that handheld devices can increase student engagement and enrich learning experiences through interactivity and visualization. To support mobile DSP laboratories, i-JDSP was developed as a visual programming environment and has been successfully tested in our undergraduate courses and high school outreach activities. i-JDSP is a graphical iOS application that enables interactive DSP laboratories on iPhones and iPads. As an example, we present in Figure 8 an i-JDSP simple digital filter simulation. i-JDSP [39] has a simple user interface associated with a minimal learning curve. DSP functions are designed as blocks that users can add to the simulation canvas. The app uses multiple views for selecting blocks and navigation is straight forward. Users open the function list, select a block, edit its parameters, and add it to the canvas. i-JDSP supported functions include: signal generator, plot, FFT, filter design, MIDI, and DTMF. All the labs described in Section II are supported by i-JDSP. i-JDSP won the Premier award [46] at FIE 2012. An Android based simulation environment was also developed and was reported in [40]. The main J-DSP web site [47] contains additional info including applications for earth systems [48]. A book [49] that contains J-DSP exercises and examples with mobile J-DSP has also been published and is used in our DSP class.

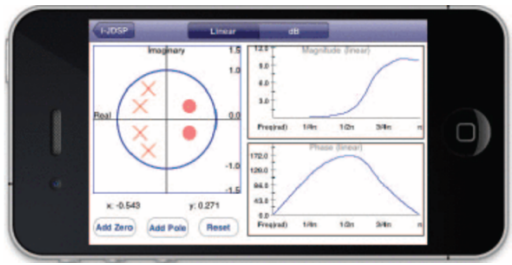

Figure 8. Mobile i-JDSP digital filter simulation. i-JDSP won the Premier courseware award [52] at IEEE FIE 2012 in Seattle.

## VII. WORKFORCE DEVELOPMENT AND OUTREACH

J-DSP has been used as a workforce development and education platform across NSF programs and high school outreach activities. It supported hands-on training in NSF REU and IRES bootcamps, providing undergraduate students with hands-on practical experiences in signal processing, machine learning, quantum computing and other related topics and applications. J-DSP has also been used in teacher training workshops [41] to help educators integrate interactive DSP computing modules into their courses. The quantum computing extensions of J-DSP, including QFT-based labs [42], were incorporated into our most recent REU program focused on quantum computing, giving teachers and community college instructors early exposure to quantum signal processing concepts. In all RET workforce programs, the teachers performed the QFT exercise and created lesson plans to deliver high level exposition with demonstrations to their students.

J-DSP was also used in international training programs and specifically in Fulbright outreach activities [44,45] in the Balkans. These international sessions included training in quantum computing and specifically on the use of the QFT to extract features for quantum machine learning. In addition, J-DSP has been used for high school outreach activities [50], including training sessions for high school students at the Hermanas [51] event hosted by the Phoenix College, helping introduce advanced STEM topics through interactive, visual laboratories. Specific functions were created for high school outreach including speech processing functions [43] that created audio effects.

## VIII. CONCLUSION

J-DSP has evolved over more than two decades from an early web-based DSP tool into a broad platform for hands-on signal processing education. It was introduced in the late 1990s and enabled online visual DSP simulations. It was integrated into the undergraduate DSP curriculum at ASU with NSF support. Online laboratories covering filters, FFTs, multirate systems, and random signals helped connect theory to practice. Assessments showed learning gains driven by interactive visualization. To address security and browser compatibility of Java applets, J-DSP was redesigned as a fully browser-based HTML5 platform. This transition improved security, performance, and browser support. The HTML5 version also enabled real-time data handling and larger-scale simulations. J-DSP functionality was expanded to support topics that bridge DSP with machine learning and quantum computing. New modules for speech processing, LPC-based feature extraction, clustering, and phoneme recognition provided data-driven AI experiences. More recent QFT and IQFT functions along with integrated noise models enabled students to compare FFT and QFT methods in signal analysis-synthesis. Beyond the classroom, J-DSP has supported NSF funded workforce development, namely REU and IRES programs, RET teacher training, and high school outreach. Mobile versions on iOS and Android were established for outreach. Dissemination and dedicated workshops engaged other universities that established mirror sites. In fact, Johns Hopkins university and MIT used the tools in earth systems and biomedical engineering courses respectively. This paper cited several of our previous publications documenting details of J-DSP related functions and hands-on activities as they evolved from 1998 to the present day. Finally, we note that the iOS version of J-DSP received the 2012 Premier courseware award presented at IEEE FIE 2012 in Seattle. In addition, a relatively recent study on J-DSP Quantum Fourier transform simulations received Top 3% recognition at IEEE ICASSP 2023.

## ACKNOWLEDGMENT

J-DSP was supported in part through the years by various NSF grants that involved undergraduate education software tool development; these grants include NSF awards 0089075, 0443137, 0817596, 1525716, and 2215998. J-DSP was also used in workforce development training associated with NSF awards 1659871, 1854273, 1953745 and 2349567.

REFERENCES

[1] Clausen, A., Spanias, **A.,** Xavier, A., and Tampi, M., "A Java signal analysis tool for signal processing experiments," *Proceedings* of *the 1998 IEEE ICASSP,* vol. **3,** pp. 1849- 1852, May 1998.

[2] Y. Cheneval, et al, "Interactive DSP education using Java," in *Proc. IEEE ICASSP* May 1998, pp. 1905–1908.

[3] J. Shaffer et al, "Visualization of signal processing concepts," in *Proc. IEEE ICASSP*, May 1998, pp. 1853–1856.

[4] The MathWorks, *The Growth of MATLAB and The MathWorks over Two Decades*, Cleve Moler, technical article, 2006. [Online]

*[5] A Kamas and E. Lee, DSP Experiments: Using a Personal Computer with Software Provided*, Prentice Hall, 1989.

[6] Spanias, A, 2001. On-Line Undergraduate Laboratories, *NSF 0089075.*

[7] Spanias, A., Urban, S., Constantinou, A., Tampi, M., Clausen, A., Zhang, X., Foutz, J. and Stylianou, G., 2000, June. Development and evaluation of a web-based signal and speech processing laboratory for distance learning. In *2000 IEEE ICASSP* (V.6, pp. 3534-3537).

[8] Spanias, A., Ahmed, K.I., Papandreou-Suppappola, A. and Zaman, M., 2003, November. Assessment of the Java-DSP (J-DSP) on-line laboratory software. In *33rd IEEE FIE 2003.* (pp. T2E-16).

[9] Spanias, A., V. Atti, A. Papandreou-Suppappola, K. Ahmed, M. Zaman, and T. Thrasyvoulou. "On-line signal processing using J-DSP." *IEEE Signal Processing Letters* 11, no. 10 (2004): 821-825.

[10] Spanias, A. and Bizuneh, F., 2001, May. Development of new functions and scripting capabilities in JavaA-DSP for easy creation and seamless integration of animated DSP simulations in Web courses. *2001 IEEE ICASSP 2001,* Vol. 5, pp. 2717-2720. IEEE.

[11] Spanias, A., Panayiotou, C., Thrasyvoulou, T. and Atti, V., 2004, June. Java Dsp Interface With MATLAB and Its Use In Engineering Education. In *2004 ASEE Annual Conference* (pp. 9-828).

[12] Spanias, A., Chilumula, R., Huang, C. Stiber, M., Loizou, P. and Kasparis, T., 2006, June. A Collaborative Project On Java Dsp Involving Five Universities. *2006 ASEE Ann. Conf. & Exposition*.

[13] K. Ramamurthy, A. Spanias, L. Hinnov, C. Akujuobi, M. Stiber, M. Pattichis, E. Doering, C. Pattichis, H. Thornburg, A. Papandreou-Suppappola, and P. Spanias, "Work in progress – collaborative multidisciplinary J-DSP software project," *IEEE FIE* San Antonio, 2009.

[14] Spanias, A., Thrasyvoulou, T., Song, Y. and Panayiotou, C., 2003, November. Using J-DSP to introduce communications and multimedia technologies to high schools. *IEEE FIE 2003,* V. 2, pp. F3A-22.

[15] Spanias, A. and Atti, V., 2005. Interactive online undergraduate laboratories using J-DSP. *IEEE Trans, on Educ.*, *48*(4), pp.735-749.

[16] Spanias, A, "A brief survey of time-and frequency-domain adaptive filters," *2016 7th IISA*, pp. 1-7. IEEE, 2016, Chalkidiki, July 2016.

[17] B. Widrow and S. D. Stearns, *Adaptive Signal Processing*. Englewood Cliffs, NJ, USA: Prentice-Hall, 1985.

[18] P. S. R. Diniz, *Adaptive Filtering: Algorithms and Practical Implementation*, 4th ed. Hoboken, NJ, USA: Wiley, 2013.

[19] L. R. Rabiner, "A tutorial on HMMs and applications in speech recognition," *Proc. IEEE*, vol. 77, no. 2, pp. 257–286, 1989.

[20] Ahmadi, S. and Spanias, "Cepstrum-based pitch detection using a new statistical V/UV classification algorithm," *IEEE Trans. on Speech and Audio Processing*, *7*(3), pp.333-338, 1999.

[21] H. Hermansky and S. B. Davis, "Toward the development of a robust connected-digit recognizer," *IEEE Trans. ASSP*, Nov. 1989.

[22] R. Gorrieri et al, "Applet Watch-Dog: a monitor controlling the execution of Java applets," *Proc. 14th (SEC'98)*, London, Sep. 1998.

[23] U. Borkowski *et al.*, "SwingJS: Giving Java applets new life as JavaScript equivalents, applications to education and science," *20. Workshop Software-Reengineering und -Evolution*, 2018, Aachen.

[24] Dixit, A., Katoch, S., Spanias, P., Banavar, M., Song, H. and Spanias, A., "Development of signal processing online labs using HTML5 and mobile platforms," *2017 IEEE FIE,* pp. 1-5. Indianapolis, Oct. 2017.

[25] Dixit, A., Shanthamallu, U., Spanias, A., Berisha, V., Banavar, M., "Online machine learning experiments in HTML5," *IEEE FIE,* 2018.

[26] Linde, Y., Buzo, A. and Gray, R., 2003. An algorithm for vector quantizer design. *IEEE Trans. Comm.*, *28*(1), pp.84-95.

[27] Shanthamallu, U.S. and Spanias, A., *Machine and Deep Learning Algorithms and Applications,* Springer International Publishing, 2022.

[28] Sasi, G. "Exploring gender classification from speech signal using Java-DSP tool." *CVR Journal of Sc. and Tech.* (2020): 28-31.

[29] Spanias, A., 2022. Quantum Machine Learning Online Materials and Software Modules for UG Education. *NSF Award 2215998*.

[30] McAulay, R. and Quatieri, T., Speech analysis synthesis based on a sinusoidal representation. *IEEE Trans. ASSP*, pp.744-754, 2003.

[31] Painter, T. and Spanias, A., 2005. Perceptual segmentation and component selection for sinusoidal representations of audio. *IEEE Transactions on Speech and Audio Processing*, *13*(2), pp.149-162.

[32] Sharma, A., Uehara, G., Narayanaswamy, V., Miller, L. and Spanias, A.," Signal analysis-synthesis using the quantum Fourier transform," *Proc. IEEE ICASSP 2023,* (pp. 1-5). Rhodes, June 2023,

[33] Sharma, A., Uehara, G., Wang, C., Larson, J., Barnard, W. and Spanias, A., "Education Software Development for Undergraduate Exposition on the Use of Quantum Fourier Transforms in Signal Analysis," *IEEE Trans. on Education, pp.* 474 – 483, *Vol. 68, Sept. 2025*.

[34] Weinstein et al, 2001. Implementation of the quantum Fourier transform. *Physical review letters*, *86*(9), p.1889.

[35] Shor, P.W., 1994, Nov. Algorithms for quantum computation: discrete logarithms and factoring. In *Proceedings 35th annual symposium on foundations of computer science* (pp. 124-134). IEEE.

[36] Behera, B.K., 2021. Simulation of Lennard-Jones Potential on a Quantum Computer. *arXiv preprint arXiv:2101.10202*.

[37] Clerk, A A., M. Devoret, S. Girvin, F. Marquardt, and R. Schoelkopf. "Introduction to quantum noise, measurement, and amplification." *Rev of Modern Physics* 82, no. 2 (2010): 1155-1208.

[38] S. Aleksandrowicz *et al*., "Qiskit Aer: High-performance quantum computing simulators with realistic noise models," *arXiv preprint* arXiv:1908.06002, 2019.

[39] Liu, J., Hu, S., Thiagarajan, J.J., Zhang, X., Ranganath, S., Banavar, M.K. and Spanias, A., 2012, Interactive DSP laboratories on mobile phones and tablets. *2012 IEEE ICASSP,* pp. 2761-2764.

[40] Ranganath, S., Thiagarajan, J., Rajan, D., Banavar, M., Spanias, A., Jaskie, K. and Tepedelenlioglu, C., 2019. Interactive Signal Processing Education Applications for the Android Platform. *Computers in Education Journal*, *10*(2).

[41] Banavar, M., Rajan, D., Strom, P., Spanias, P., Zhang, X, Braun, H. and Spanias, A., 2014, October. Embedding Android signal processing apps in a high school math class—An RET project. In *2014 IEEE FIE*.

[42] T. Patel, J. Larson, G. Uehara, N. Babar, D. Pujara, A. Spanias, "Training Students for Research with Quantum AI Simulation Tools," Accepted at *IEEE ICASSP-26,* Barcelona, May 2026.

[43] A.S. Spanias, "Speech Coding: A Tutorial Review," *Proceedings of the IEEE*, Vol. 82, No. 10, pp. 1441-1582, October 1994

[44] A. Spanias, "International Research and Workforce Development in Machine Learning for Energy Applications - A Fulbright US Scholar Project," *IEEE 16 th IISA 2025*, Mytilene, July 2025.

[45] G. Uehara, E. Vasiileva, M. Chausevska, M. Todevska, J. Kamchev, Z. Ivanovski, D. Dimitrov, A. Spanias, "International Student Training in Quantum Machine Learning," *IEEE 16th IISA 2025*, Greece, July 2025.

[46] Premier Award https://jdsp.engineering.asu.edu/award2012/index.html

[47] Java Digital Signal Processing web site. http://jdsp.asu.edu

[48] Ramamurthy K., Spanias A., Hinnov L., Ogg J., "On the use of Java-DSP in Earth systems", *Proc. of ASEE*, Pittsburgh, June 2008.

[49] A. Spanias, Digital Signal Processing; An Interactive Approach, 2nd Edition, ISBN 978-1-4675-9892-7, Lulu Press, Morrisville, May 2014.

[50] Outreach CDS https://jdsp.engineering.asu.edu/CDS_High_AJDSP/

[51] Hermanas Conf https://jdsp.engineering.asu.edu/Hermanas_Conference/

[52] https://www.premieraward.org/premier-courseware-of-2012/index.html

[53] https://2023.ieeeicassp.org/top-3-percent-paper-recognitions/